\documentclass[a4paper,fleqn,usenatbib]{mnras}
\usepackage{newtxtext,newtxmath}

\usepackage[T1]{fontenc}
\usepackage{ae,aecompl}

\usepackage{graphicx}	
\usepackage{amsmath}	
\usepackage{amssymb}	

\usepackage{subfig}
\title[Black holes and light element variations in GCs]{Light element variations in globular clusters via nucleosynthesis in  black hole accretion discs}

\author[Breen]{
Philip G. Breen$^{1}$\thanks{E-mail: phil.breen@ed.ac.uk}
\\
$^{1}$School of Mathematics and Maxwell Institute for Mathematical Sciences, University of Edinburgh, Kings Buildings, Edinburgh EH9 3JZ\\
}

\date{Accepted XXX. Received YYY; in original form ZZZ}

\pubyear{2015}

\begin{document}
\label{firstpage}
\pagerange{\pageref{firstpage}--\pageref{lastpage}}
\maketitle

\begin{abstract}
Ancient globular clusters contain multiple stellar populations identified by variations in light elements (e.g., C, N, O, Na). Although many scenarios have been suggested to explain this phenomenon, all are faced with challenges when compared with all the observational evidence. In this Letter, we propose a new scenario in which light element variations originate from  nucleosynthesis in accretion discs around black holes. Since the black holes form after a few $Myrs$, the cluster is expected to still be embedded in a gas rich environment. Through a simplified accretion model, we show that the correct light element anti-correlations can be produced.
Assuming a Kroupa stellar initial mass function (IMF), each black hole would only have to process ${\approx}300M_{\odot}$ of material in order to explain multiple populations; over a period of $3Myr$ this corresponds to $ \sim10^{-4} M_{\odot}yr^{-1}$ (similar to the estimated accretion rate for the x-ray binary SS 433). 
\end{abstract}




\section{Introduction}

Ancient globular clusters have been found to exhibit light element variations in the form of anticorrelation in $Na-O$ and $N-C$, where some stars show enrichment in $Na$ ($N$) and depletion in $O$ ($C$)  relative to field stars with the same metallicity \citep[see][for a review]{GCB2012,BL2017}. Evidence has also been found of multiple populations in young massive clusters (YMC), with ages in the range 2-8 Gyr, see \cite{K2016} for a review. 
So far no evidence of multiple populations have been found in massive clusters younger than $\sim$2 Gyr \citep[e.g.,][]{Mu2008,Mu2014,MB2017}.

A number of scenarios have been put forward to explain this phenomenon, usually involving pollution by a first generation of stars. The most studied scenario involves pollution by slow moving winds from AGB stars \citep[see e.g.][]{CD1981,D2008,B2017}, which cools and collects in the center of the cluster to form the polluted stars. However, this scenario requires clusters to have an initial mass, several times the present value, the so-called mass budget problem \citep{D2008,C2012}. 


Other possible sources of pollution include fast rotating massive stars \citep{D2007},  interacting binaries \citep{D2009}, stripped envelopes of high mass stars \citep{PC2006,E2017} and a single very massive star \citep[][]{DH2014,Gi2018}. It has also been suggested by \cite{M2009} that the order in which the populations form may be reversed; this scenario involves a pre-enrichment phase with Type II supernovae as well as local enrichment by a single Type Ia SN and AGB stars. Another possibility is that the enriched stars are not the result of multiple epochs of star-formation but the result of the pollution of protoplanetary discs of low mass stars \citep{B2013}. 

All current scenarios face challenges when compared to all the observational constraints \citep[e.g., see Fig. 6 in ][]{BL2017}. One challenge for all models is the ability to create discrete sub-populations \citep[e.g.][]{Mil2015}; there is also a need for stochasticity in the pollution process in order to explain the high degree of cluster-to-cluster variations  \citep{B2015,Mil2015}. Another constraint is that properties of the pollution mechanism seem to be correlated to the cluster mass; for example both the fraction of polluted stars and the size of the abundance spread is correlated with the mass of the cluster \citep{M2017}.  {  Evidence that YMC have been found to be gas free at $\sim7Myr$ or earlier, appears to indicate that multiple populations are at least approximately coeval  \citep[see][and references therein]{BL2017}, if clusters are required to retain pristine gas in order to produce multiple populations.}

Here, we propose that light element variations can be generated in accretion discs around stellar-mass black holes.

\section{Black hole accretion scenario}
\label{sec:bham}
\subsection{Stellar mass black holes in globular clusters}
\label{sec:bhgc}
Assuming a \cite{Kroupa2001} IMF,  the total number of black holes ($N_{bh}$) expected to form is $ N_{bh} \approx   2.2\times10^{-3} N_{*}$, where $N_{*}$ is the total number of stars in the cluster (assuming a mass range of $0.1-100.0M_{\odot}$ and that all stars over $25M_{\odot}$ become black holes). Some black holes might escape the cluster shortly after formation due to a ``natal kick'' resulting from asymmetries in the supernova; however, if the most massive black holes (with progenitor mass $\gtrsim 40M_{\odot}$) form via direct collapse then there is no supernova and no ``natal kick" \citep[e.g. see][]{F1999,FK2001,H2003,F2012}, therefore these objects would be retained in the cluster. The fraction among black holes with progenitors in the mass range $40-100M_{\odot}$ is approximately  $0.45$. \cite{AAG2018} preformed Monte Carlo simulations of globular clusters using a natal kick prescription from \cite{Bel2002} and found at 20-30Myr the black hole retention fraction ($f_{r,bh}$) is in the range $0.15$ to $0.55$. 

The present day black hole population of globular clusters is poorly constrained; a number of black hole candidates (in binaries with luminous stars) have been identified in globular clusters \citep{Mac2007,Str2012,Cho2013,G2018}. \cite{AAG2018b} proposed a novel way to identify the size of the black hole population, using a concept similar to ``influence radius'' and found that many of the Milky Way's globular clusters likely host as many as several hundreds of black holes, at the present time. Note that it is the initial population that is important here and not the present-day one, as over time the black hole population decreases due to dynamical ejection \citep[see e.g.][]{BH2013}.

\subsection{Total amount of enriched gas}
\label{sec:TEG}
Lithium is destroyed at temperatures well below those needed to produce the $Na-O$ anti-correlation. However enriched stars have been found to contain Lithium abundance comparable to pristine stars \citep[see e.g.][]{Mu2011}. Therefore enriched stars consist of a mixture of pristine and enriched gas \citep{PC2006}. We will assume that the diliution fraction, i.e. the fraction of enriched gas per star ($f_d$), is $~0.5$. The present day fraction of enriched stars ($f_s$) ranges from $~0.33$ to $~0.92$. However the fraction is likely to have increased with time as pristine stars are preferentially stripped from the outskirts of the system. If we assume that the initial stellar mass of a globular cluster ($M_*$) is $4.5$ times the present day mass \citep{WL2015} then the initial range of $f_s$ is $~0.07$ to  $~0.20$. We will adopt a value at the top of this range $f_s = 0.2$. Finally we will consider the star formation efficiency of the enriched gas ($\epsilon_{sf}$), i.e. the amount of enriched gas that ends up in stars.  We will assume that $\epsilon_{sf}=0.4$, as found by \cite{Bon2011} when the material is maximally bound. The total mass in enriched gas ($M_g$) is therefore: 
$$ M_g =    \epsilon_{sf}^{-1} f_{d}  f_{s} M_{*} \approx 0.25 M_{*},$$
where $M_*$ is the initial stellar mass of the cluster. We adopt a black hole retention fraction of $f_{r,bh}=0.35$, in the middle of this range given by \cite{AAG2018}, i.e. , each black hole would then only be required to process: $$ \frac{M_g}{f_{r,bh}N_{bh}} = \frac{ 0.25M_{*} }{   2.2\times10^{-3} f_{r,bh} N_{*}}  \approx 300  M_{\odot}.$$

\subsection{Accretion, outflows and feedback}
\label{sec:AOF}
Since the black holes are formed after a few $Myr$, we assume that the conditions are similar to when star formation began, i.e. the environment is gas rich, consisting of a hierarchical structure with many dense clumps \citep{MO2007,T2014}. As in the Competitive Accretion model of massive star formation \citep{Bonnel2001}, we argue that the potential of the clusters acts to funnel gas down to the center of the cluster creating a gas reservoir which can be replenished by gas inflow. If we assume that the black holes are in clumps of total mass $10^3 M_{\odot}$ and radius $0.1$pc than the typical velocity would be $v \sim \sqrt{GM/r} \sim  7$km/s, using $M_{BH} = 10M_{\odot}$ and $\rho_g = 10^{-16} g cm^{-3}$ then the resulting accretion rate 
is $\dot{M}_{ac} = 4\pi \rho_g (GM_{BH})^2/v^3 \sim10^{-4}M_{\odot}yr^{-1}$, at these rates the enriched gas could be produced in $\approx 3Myr $. { Our crude estimate places the model just within the constraints from YMC being gas free by $\sim7Myr$ and shorter for a lower fraction of enriched stars.} Note the accretion rates based on Bondi-Hoyle formalism can be very sensitive to small changes in the physical parameters \citep{BonnelB2006}, and therefore the accretion rate is highly uncertain.  


For a $10M_{\odot}$ black hole, the Eddington luminosity ($L_{ed}$) is reached at an mass accretion rate of $\approx 10^{-8} M_{\odot}yr^{-1}$, assuming a radiative efficiency of $0.1$, i.e. $\dot{M}_{ed}=L_{ed}/(0.1c^2)$. The accretion rates considered above are orders of magnitude above Eddington, so it is expected that most of the accredited mass escapes in outflows \citep[e.g., see][]{KP2003}. Our model requires that enough of the gas is exposed to high temperatures before escaping. \cite{SS1973} suggested that most of the mass would be lost from the radius at which $L_{ed}$ is first reached. A simple model for the outflow assumes $\dot{M} \propto r^s$,  where $s$ is in the range $0 < s < 1$ \citep{BB1999}. Using $s=1$ \citep{Be2012}, the outflow from the region where our simple model predicts the light element enrichment are generated (see Sec \ref{sec:CAD} \& \ref{sec:ny}) is approximately $70\%$ of the inflow. 



Using ultra-luminosity X-ray sources ULXs as a guide \citep[e.g., see][]{Kaa2017} the outflows could have speeds of  $\sim 100$ to  $\sim 1000km s^{-1}$.  These would be in excess of the escape velocity. However, as long as the cluster is embedded the outflows will loose kinetic energy when they sweep up pristine gas. \cite{WTE2017} considered fast moving winds with radiative cooling of gas and found a significant fraction of fast winds ($\gtrsim 1000kms^{-1}$) could be retained within the cluster. { Even if the the bulk of the black holes form without supernova (via direct collapse), there may still be some pair instability supernovae and supernova from stars with $\leq 40M_{\odot}$, however, as winds from supernova travel an order of magnitude faster ($\sim10^4$ km/s) they may preferentially escape with little effect on the gas.}

If the black holes were all radiating at the Eddington Luminosity, since the mass budget in the central region will be dominated by gas,the overall gravitational attraction will far exceed the radiation pressure.  Over time the radiation and winds from the black holes will heat the gas and eventually deplete it, estimates from \cite{Leigh2013} require $\gtrsim 10Myr$ for gas expulsion which will allow sufficient time for the enriched stars to form.

\subsection{Conditions in the accretion disc}
\label{sec:CAD}
The topic of accretion discs around black holes is a rich and interesting subject; see \cite{A2013} for a review on the subject. One of the most studied accretion disc models is the thin disc model of \cite{SS1973}, where the model is geometrically thin, optically thick, radiatively efficient and uses a dimensionless constant, $\alpha$, to parameterize uncertainty over the viscosity mechanism. The model also admits analytic solutions under assumptions of the pressure and opacity \citep[for the general relativistic versions see][]{NT1973}. However, as we are concerned with high accretion rates the thin disc model, is unsuitable because the assumption that the disc is radiatively efficient breaks downs as cooling by advection becomes more important \citep[e.g. see][]{A2013}.

\begin{figure}
\includegraphics[ trim=0.7cm 1.3cm 1.0cm 1.cm, clip, width=\columnwidth]{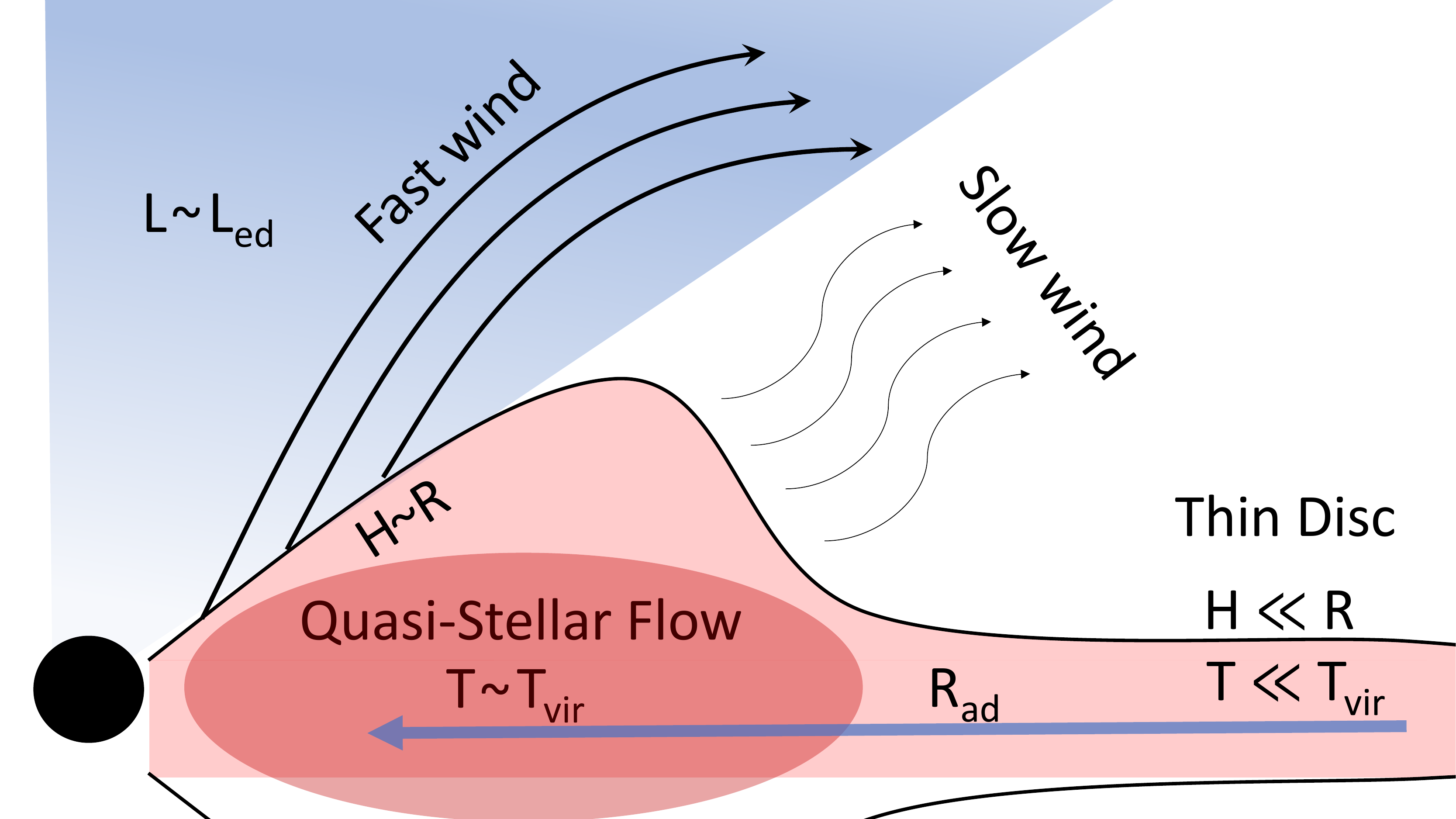}
\caption{Schematic of accretion flow around the black hole. The blue arrows gives the direction of the flow. Far from the black hole, where advection is unimportant and radiative cooling is efficient the conditions are similar to that of a thin disc. Here the height of the disc ($H$) is small compared to the radius ($R$) and the temperature $(T)$ is low compared to the virial temperature. At the radius at which advection becomes important ($R_{ad}$), radiative cooling becomes inefficient and heat generated by viscosity is advected with the flow, causing the flow to heat and expand. This results in the flow becoming quasi-spherical (i.e. $H \sim R$ ) and temperatures becoming approximately virial (assuming cooling is inefficient).  This region has been labeled ``Quasi-Stellar Flow", as we assume that conditions here are similar to stellar interiors. The flow is constantly losing mass in the form of winds which increase in speed with decreasing $R$, resulting in the accretion rate dropping to $~\dot{M}_{ed}$ closer to the black hole.
}
\label{fig:QSF}
\end{figure}

In advection-dominated accretion flows (ADAF) temperatures can reach almost virial values \citep{R1982,NY1994,NM2008}, though they are usually studied in the optically thin regime where flow is radiatively inefficient due to low densities. At the high accretion rates considered in Sec \ref{sec:AOF} the disc would be both geometrically thick and optically thick. If the optical depth is large enough the radiative cooling becomes inefficient and more of the energy generated by viscous heating would be retained and advected in the flow. Note that in a radiatively inefficient flow the gravitational potential energy cannot be radiated away. Once advection becomes important, the accretion flow will increase in temperature reaching approximately virial values (assuming cooling mechanisms are inefficient). Since the conditions may become similar to stellar interiors, the term ``Quasi-Stellar Flow" seems appropriate. A schematic representation is given in Fig. \ref{fig:QSF}.

Also, the maximum possible temperatures reached in a steady state flow around a black hole are approximately virial, as temperatures in excess of this would result in expansion. In order to estimate the light element yields, we assume that an approximately virial temperature profile has been reached: $$t_{vir} = \frac{GM_{bh}m_p}{6kr} \approx 3.8\times10^{7} \bigg( \frac{M}{10M_{\odot}}  \bigg)  \bigg( \frac{r}{R_{\odot}}  \bigg)^{-1} K,$$ 
where $R_{\odot}$ is a solar radius. Note that even at approximately virial temperatures the flow could still have appreciable radial velocities.

\subsection{Nucleosynthesis in accretion discs}
\label{sec:ny}

Nucleosynthesis usually is not included in models of accretion discs, though  there has been some work on the topic.  \cite{MC2000} showed that significant nucleosynthesis could take place around a $10M_{\odot}$ and a $10^6M_{\odot}$ black hole if the viscosity is low enough \cite[also see][]{AH1992,HP2008}.

The $Na-O$ and $N-C$ anti-correlations are produced at roughly $5 \times 10^7K$, which under the assumption of a virial temperature profile (see Sec \ref{sec:CAD}), corresponds to a radius of $ \approx 5.4\times 10^{10}cm$ ($\approx 2\times10^4R_s$). We will consider a shell from $T=2.5 \times 10^7 - 7.5 \times 10^7 K$ and assume that yields will be similar to that of the composition being evolved at fixed $\rho$ and $T=5\times 10^7 K$. The yields were calculated using the publicly available nuclear reaction network Torch by \cite{T1999}\footnote{\url{http://cococubed.asu.edu/code_pages/burn.shtml}} and the initial composition used is the same as that given in \cite{D2007} (their Table 3) corresponding to a metallicity of $[Fe/H]=-1.5$. The initial and final mass fractions of  C, N, O and Na are given in Table \ref{tab:ele}. Radial velocities ($v_r$) are calculated by dividing the the width of a shell by time spent at the temperature.

\begin{figure}
	\subfloat[]{{\includegraphics[trim=0.0cm 1.cm 1.4cm 0.7cm, width=\columnwidth]{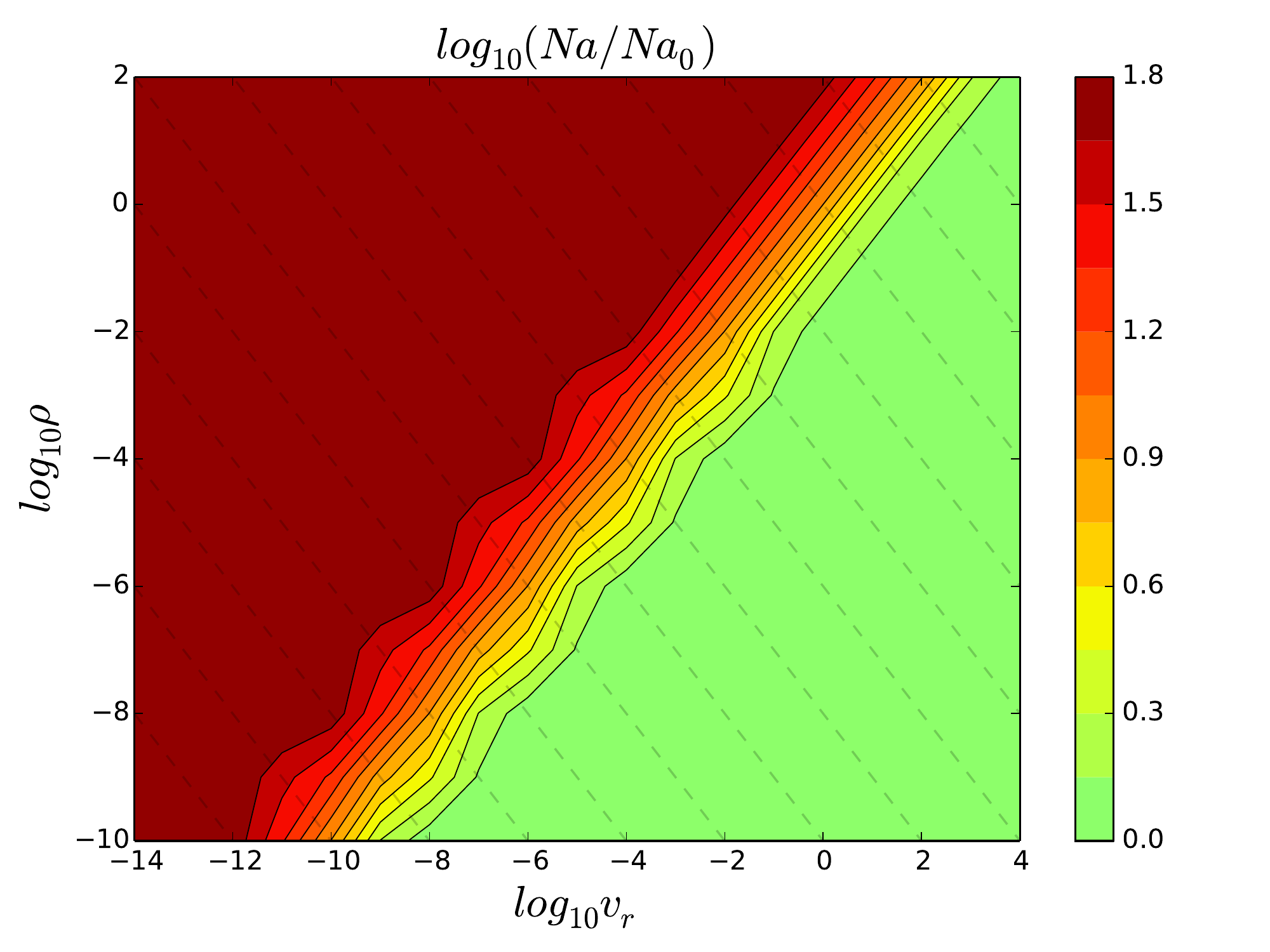} }}%
    \caption{ Ratio of initial to final Na as a function of mean $\rho$ and $v_r$.  For reference the free fall velocity in this region is $v_{ff} \sim 10^8 cm/s$. The dashed lines correspond to fixed accretion rates assuming quasi-spherical accretion i.e. $\rho \approx \dot{M}/(4\pi r^2 v_r)$. 
    For a range of values Na is enriched by $10^{1.8}$, so long as $v_r$ is sufficiently low or the $\rho$ is sufficiently high, see text for details. }
\label{fig:temp}
\end{figure}

The results for Na enrichment are shown in Fig. \ref{fig:temp}. As long as the flow is slow enough, for a given mean density, $Na$ is significantly enriched by a factor $58$ (and similarly $N$ by  $30$). Assuming a dilution factor of $f_{d}  = 0.5$ (see Sec \ref{sec:TEG}), this will produce a spread of $1.5$ dexs in $Na$  ($1.1$ for $N$) between enriched and pristine stars. Similarly $O$  and $C$ are reduced by factors $0.01$ and $0.1$, since these values are small, the spreads in $O$  and $C$ result from the dilution of pristine gas and will be approximately $log_{10}(1-f_d)$.

\begin{table}
\caption{ $\alpha$ element symbol, $M_{fr,i}$ initial mass fraction,  $M_{fr,f}$ final mass fraction and $\log_{10}(M_{fr,f}/M_{fr,i})$  enrichment factor.}\label{tab:ele}
 \begin{tabular}{|| c c c c  ||} 
 \hline 
   $\alpha$ \quad &  $M_{fr,i}$  & $M_{fr,f}$  & $\log_{10}(M_{fr,f}/M_{fr,i})$ \\ [1.0ex] 
 \hline\hline
  C & $3.50 \times 10^{-5}$  &   $3.39 \times 10^{-6}$  & -1.01 \\ 
  N & $1.03 \times 10^{-5}$  &   $3.13 \times 10^{-4}$  & 1.48 \\ 
  O & $3.00 \times 10^{-4}$  &   $3.02 \times 10^{-6}$  & -2.00 \\ 
  Na & $3.30 \times 10^{-7}$  &   $1.94 \times 10^{-5}$  & 1.76 \\ 
\end{tabular}
\end{table}

\section{Discussion}
We have argued that the light element variations observed in globular clusters could be generated in accretion discs around stellar-mass black holes. In this scenario the black holes form from the earliest massive stars after a few $\sim Myr$ and are born into a gas rich environment. The black holes are expected to experience similar accretion rates as those estimated for the formation of a massive star. The accreted gas is then subjected to high temperatures in the inner part of the disc after which it escapes in outflows. If temperatures in the accretion disc are high enough significant nucleosynthesis will occur and the required light element variations can be generated. The polluted material escapes from the disc in outflows which mix with pristine gas, from which the enriched stars are formed. 

Using a highly simplified model we have shown that a population of stellar-mass black holes could produce sufficient material to produce a polluted population of stars with $Na-O$ and $N-C$ anticorrelation, which is the main signature of multiple populations. These results need to be compared with more detailed models and numerical simulations of accretion flow, but the aim of this Letter is to show that such a model is plausible and warrants further study.  

The model presented in this Letter has many promising features. First, since pollution takes place over a short time scale $\lesssim10$Myr the populations are co-evolved. Second, since the black holes process the pristine gas, there is no mass budget problem and there is also a natural dilution mechanism as outflows from the accretion disc sweep up pristine gas. Third, since accretion rates and times depend on the properties of the host system, the fraction of enriched stars and abundances are expected to correlate with the cluster mass. Fourth, it has been recently suggested by \cite{C2018} that simple dilution models cannot explain the light element variations observed in NGC 2808 and that polluters of different masses are required. This might not be an issue for the model proposed because the black holes will have a range of masses and even for a single accretion disc the composition of the wind (and wind speed) will likely vary with radius in the  accretion disc. Fifth, since higher temperatures than those considered in Sec \ref{sec:ny} are expected at smaller radii, the scenario could potentially explain small Fe spreads measured in some globular clusters \citep[e.g. M22][]{DaC2009, Mar2009, Mar2011} without the need to retain supernova ejecta. Interestingly, if the enriched stars are formed in the vicinity of accreting black holes then their stellar masses may be limited by competitive accretion \citep{Bonnel2001}. If the maximum mass is limited to below the main sequence turn-off for a cluster of $\sim 2Gyr$, then this would explain why multiple populations  have not been observed in YMC below this age  \citep[e.g., see][]{BL2017}.

%
%

In this letter we have made the assumption that cooling is inefficient in the accretion flow. If the optical depth is large it is reasonable to assume that radiative cooling is inefficient, however, it is possible the flow may cool by another mechanism e.g. energy loss through outflows. A key next step to develop the scenario, is to construct a more detailed model of the accretion flow. This will help constrain the conditions required to reach the temperatures needed for nucleosynthesis to occur. We have also made the assumption that the black holes will experience similar accretion rates as for massive star formation. However, the feedback from black holes will be different which may affect the accretion rates.  Also feedback from the black holes may effect the star formation rates. Hydrodynamic models are required to further explore these issues and will be one of the next steps in developing the model.


The greatest uncertainty in the scenario is whether nucleosynthesis does actually occur in accretion flows around black holes. A good candidate to test the possibility of nucleosynthesis in accretion flows is SS 433 \citep[for review see][]{F2004}, which is believed to be accreting at similar rates as considered in the letter. There is already evidence of an overabundance of elements \citep[Ni, Si and S, see][]{BKK2005} in the jets of SS 433. These elements have heavier atomic numbers than $Na$, in the scenario only a small percentage ($\sim 0.01 \%$) of the overall accretion flow makes it to the jet and this material would be exposed to higher temperatures than the bulk of the inflow. A prediction of the model presented in this Letter, is that the outflows of SS 433 will be enriched in light elements.

If we consider a cluster of $N_*=10^6$ stars and if all the black holes are radiating at the Eddington luminosity, then the luminosity would be $\sim 10^{42} $ erg/s, i.e. at the top end of the range for an ultra-luminosity X-ray source (ULX) or the lower end of an active galactic nucleus (AGN). However, embedded clusters are likely Compton thick and obscured in X-ray \citep[also see][]{K2012}, except during the brief period when most of the cluster gas has been expelled and before the accretion discs have been exhausted. If a black hole is dynamically ejected with its accretion disc intact, then it will become visible once free of the cluster and may appear as an ULX, if the emission of radiation is anisotropic \citep{K2001}. Indeed a physical association between the ULXs and super star clusters in large starburst galaxies has been suggested \citep[e.g. see][]{Kaa2004,P2013}. However, the nature of ULXs is still an open question and most probably consists of a class of objects  \citep[for a review see][]{Kaa2017}.

In the scenario, we have only considered stellar mass black holes. Though it is possible that some enrichment could have been produced by an intermediate mass black hole or even a  super-massive black hole. In this context, the origin of the large population of N-rich stars in the bulge{\textbackslash}inner halo \citep{S2017} would have originated from pollution by our galaxies supermassive black hole.


\section*{Acknowledgements}
The author is grateful to Anna Lisa Varri, Douglas Heggie, Glenn van de Ven and Enrico Vesperini for many helpful comments on a draft version of the manuscript. The author would also like to thank Maximilian Ruffert and Ken Rice for interesting scientific discussions, { and the anonymous referees for their many helpful comments.} The author is very grateful to Frank Timmes for his publicly available nuclear reaction network codes. The author acknowledges support from the Leverhulme Trust (Research Project Grant, RPG-2015-408)







\appendix




\bsp	
\label{lastpage}
\end{document}